\documentclass[twocolumn]{aa}
\usepackage{graphicx}
\begin{document}
\title{Self-bound CFL stars in binary systems : are they
``hidden'' among the black hole candidates ?}

\author{J. E. Horvath \inst{1} \and G. Lugones \inst{2}}
\institute{Instituto de Astronomia, Geof\'\i sica e Ciencias Atmosf\'ericas\\
Rua do Mat\~ao 1226, 05508-900 S\~ao Paulo SP, Brazil\\
\email{foton@astro.iag.usp.br} \and
Dipartimento di Fisica ``Enrico Fermi'', Universit\`a di Pisa\\
via Buonarroti 2, I-56127 Pisa, Italy\\
\email{lugones@df.unipi.it}}

\abstract{The identification of black holes is one of the most
important tasks of modern astrophysics. Candidates have been
selected among binary stars based on a high mass function, and
seriously considered when the lower mass limit exceeds $\sim \, 3
\, M_{\odot}$. More recently the absence of (Type I) thermonuclear
bursts has been advanced as an additional criterion in favor of
the black hole interpretation, since the absence of a solid
surface naturally precludes the accumulation and ignition of
accreting material. We discuss in this {\it Letter} the
possibility that self-bound stars made of CFL-paired quarks mimic
the behavior of at least the low-mass end black holes as a result
of: a) higher maximum masses than ordinary neutron stars, b) low
steady luminosities due to the bare surface properties, and c)
impossibility of generating Type I bursts because of the complete
absence of normal matter crusts at their surfaces. These features
caution against a positive identification of event horizons based
on the lack of bursts.

}

\date{Received February 2004}

\titlerunning{Self-bound CFL stars in binary systems}
\maketitle

\section{Introduction}

The strong gravitational field regime of General Relativity is
widely believed to be physically realized in black holes, uniquely
possessing event horizons. The positive identification of black
holes among existing astrophysical objects has the potential of
providing a glimpse into that regime which captures the
imagination of scientists and public alike. This is why the study
of black hole candidates has consumed so much time of research
efforts since the early days of X-ray astronomy, when the first
strong sources were identified and some of them tentatively
associated with astrophysical black holes.

Unfortunately, the task of a positive identification is still
plagued with caveats, and even though the advances in the last
decade or so have been quite impressive (\cite{Orosz}), the burden
of proof is still with observational astrophysicists. Meanwhile,
the discussion continues and new proposals arise as potential
unique signatures of event horizons possessed by black holes only.
In a recent series of papers (\cite{Nar1,Nar2} and references
therein), Narayan and coworkers have proposed a class of X-ray
binaries (the soft-X transients, or SXT) as attractive candidates
to black holes. The argument is that those sources show, in
addition to a mass function $f(M)$ larger than the expected $\sim
\, 3 \, M_{\odot}$ maximum limit for NS models (the mass function
is an absolute lower limit to the mass of the compact object in
the SXT), an absence of type I bursts tentatively interpreted as
evidence for an event horizon. In fact, a systematic study
undertaken (\cite{Nar2}) to pin down the physical condition for
type I thermonuclear bursts has shown that SXT sources should
burst if they possess a normal matter crust, therefore the absence
of bursts could be interpreted as indicating the presence of event
horizons. In a recent paper, Yuan, Narayan and Rees (2004) made a
general attempt to show that {\it any} neutron star (composed by
matter described by a more or less general equation of state)
should experience thermonuclear Type I bursts at appropriate mass
accretion rates. The question is whether an ``abnormal'' surface
also allows such physical behavior. We shall argue below that
general models do not necessarily possess the anomalous
high-density surface properties of self-bound quark stars, which
is one of the key ingredients that entangle their discrimination
within the black hole candidates.

Since the proof of the existence of event horizons is so important
for modern astrophysics, careful examinations of the possible
loopholes and alternatives to the signatures are needed. As an
example of the latter, a critical analysis by Abramowicz,
Klu\'zniak and Lasota (2003) concluded, based on a series of
persuasive arguments, that a positive proof of the event horizon
will be impossible in principle. Specifically, they argue that the
case of SXTs as black holes based on the absence of type I bursts
would be weakened by the finding of any kind of exotic star
without a ``normal'' crust. We have reexamined this objection with
the aim of addressing the recently proposed models of self-bound
CFL stars and report our findings below.

\section{Stability of color flavor locked (CFL) quark matter and stellar sequences}

It is widely accepted that at high temperatures and/or high baryon
number densities confinement of quarks will be lost. The phase
diagram of QCD seems to be very rich and lattice simulations
(\cite{Fodor,Kanaya}) are attempting to address this point in full
detail. However, it is still not very clear whether the transition
points are of actual interest in the low temperature/high density
regime reached inside existing compact stars. Different
compositions of deconfined matter have been analyzed in the
literature, and it seems that at high densities a color-flavor
locked state of $u, d$ and $s$ quarks is favored over other forms
of paired and unpaired matter on general grounds
(\cite{Alford1999,ALF}).

We have previously addressed the issue of CFL matter absolute
stability. The idea is an extension of the celebrated ``strange
matter'' case, and in fact, we found (\cite{LH1}) that pairing
{\it enhances} the possibility of absolute stability. The CFL
phase at zero temperature has been modelled as an electrically
neutral and colorless gas of quark Cooper pairs, in which quarks
are paired in such a way that all the flavors have the same Fermi
momentum, and hence the same number density (\cite{rajw}). The
model allows CFL strange matter to be the true ground state of
strong interactions for a wide range of the parameters $B$, $m_s$
and $\Delta$, and this is why we have called this state ``CFL
strange matter'' (\cite{LH1}). The exploration of the stellar
sequences (see also \cite{LH2,AR} for a discussion of the physics
and stellar models) constructed with the values of the parameter
space of the EOS that give absolutely stable CFL strange matter,
yielded very compact configurations (which could help explain the
recently claimed compactness of a few neutron stars). In addition,
for some parameter choices the models are quite massive and
extended (see \cite{LH2}).

\begin{figure}
\includegraphics[angle=-90,width=9cm,clip]{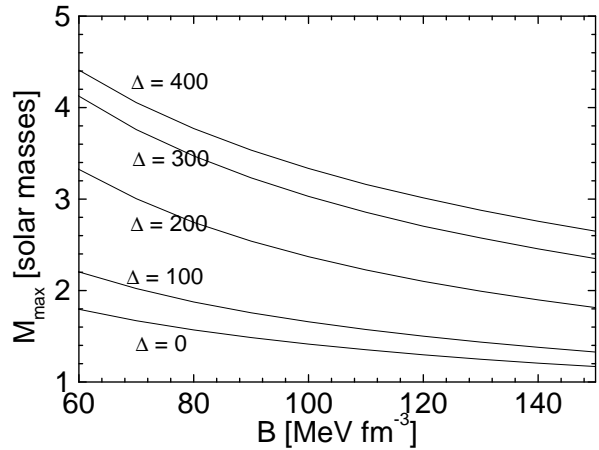}
\caption{The maximum masses of self-bound CFL stars as a function
of the bag constant $B$, for different values of the pairing
energy $\Delta$. The strange quark mass is set to $m_s = 150$ MeV.
In order to be absolutely stable the energy per baryon of the CFL
phase must be lower than 939 MeV  and the bag constant must be
greater than 57 MeV fm$^{-3}$. The stellar models were constructed
using the equations of state derived in \cite{LH1}. Large maximum
masses are found for ``standard'' values of $B$ and $\Delta \,
\geq \, 250 \, MeV$.}
\end{figure}

\begin{figure}
\includegraphics[angle=-90,width=9cm,clip]{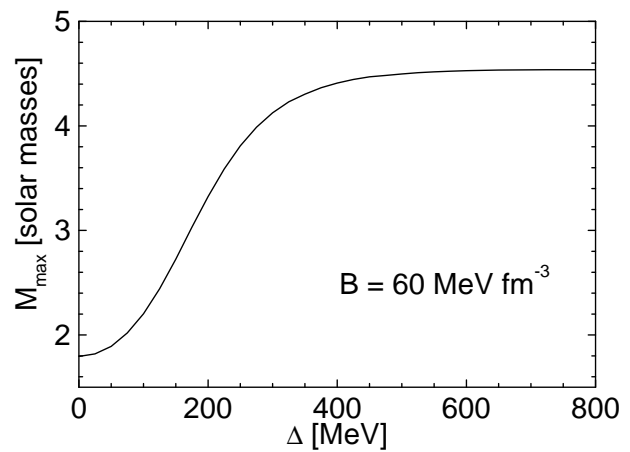}
\caption{The maximum masses of self-bound CFL stars as a function
of the pairing gap $\Delta$ for  $B=60$ MeV fm$^{-3}$. }
\end{figure}

\begin{figure}
\includegraphics[angle=-90,width=9cm,clip]{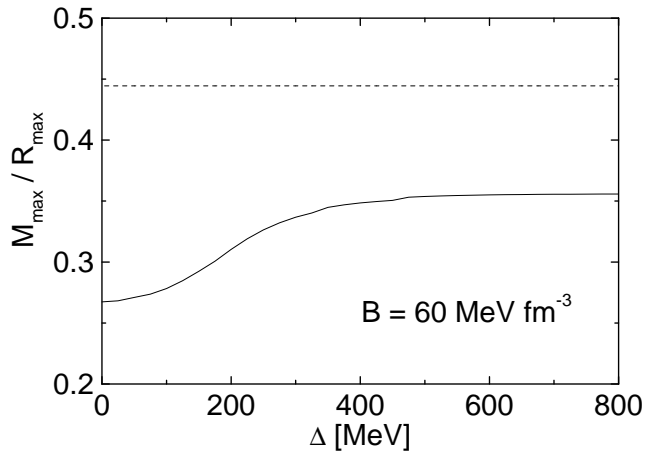}
\caption{The compactness $G M_{\rm max} / (c^2 R_{\rm max})$ as a
function of the pairing gap $\Delta$ for $B=60$ MeV \, fm$^{-3}$.
In dashed line it is shown the maximum limit $M/R = 4/9$ imposed
by general relativity (i.e. for a uniform density star with the
causal equation of state $P= \varepsilon$). Note that the
gravitational redshift of the maximum mass configuration is
$z_{\rm max} = [1 - 2GM_{\rm max}/(c^2 R_{\rm max})]^{-1/2}-1
\approx G M_{\rm max} / (c^2 R_{\rm max})$ and has an asymptotic
value $\sim 0.35$. }
\end{figure}

With the aim of addressing the SXT systems we have explored the
parameter space to check whether self-bound CFL stars sequences
end with very massive stars. Actually, the basics of $M_{\rm max}$
for any fluid configuration has been discussed by \cite{Haen} in a
pedagogical form. That work highlighted the value of scaling laws
for gaining insight of the problem, particularly for the case of
linear equations of state that allow simple solutions of the
stellar structure. In the case of strange quark matter, if
quark-quark interactions are ignored and the mass of the $s$ quark
is small enough,  the maximum mass of a stellar sequence would be
just (Haensel 2003)

\begin{equation}
M_{\rm max} = \frac{1.96 M_{\odot}}{\sqrt{B_{60}}}
\end{equation}

\noindent (being $B_{60} \equiv B / [60 \, {\rm MeV fm^{-3}}]$) as
already discussed by \cite{Witten}.

We note that the case of a CFL equation of state in fact closely
resembles the well-known theory of strange quark matter, having an
equation of state of the form $P = \kappa (\rho - 4 B_{\rm eff})$,
with $B_{\rm eff} = B - 3 \mu^2 \Delta^2 / \pi^2$
(\cite{LH1,LH2}). This parametrization, however, yields a $B_{\rm
eff}$ which is dependent on density through the chemical potential
$\mu$, and therefore a simple scaling of the maximum mass of the
star is not possible as in the case of strange matter. However,
and in order to gain a qualitative insight of the structural
properties, we can approximate the chemical potential in $B_{\rm
eff}$ by a characteristic mean value $\mu_* \sim 200 MeV$ for CFL
strange matter. In this simplified case $B_{\rm eff}$ is a
constant that depends just on $B$ and $\Delta$ and allows a very
simple scaling

\begin{equation}
M_{\rm max} \approx \frac{1.96 M_{\odot}}{\sqrt{B_{60}}} \; (1 +
\delta)
\end{equation}

\noindent valid for $B > 3 \mu_*^2 \Delta^2 / (2\pi^2)$, where

\begin{equation}
\delta = 0.15 \bigg( \frac{\Delta}{100 \; {\rm MeV}}
 \bigg)^2 \bigg( \frac{60 \, {\rm MeV fm^{-3}}}{B} \bigg).
\end{equation}

\noindent Therefore, and ignoring for the moment the entanglement
between the quark masses and the density dependence of the
condensation term, we may state that CFL matter can produce high
mass compact stars because the existence of pairing reduces the
effective value of the vacuum energy density parametrized by $B$
by adding a term with opposite sign. It is this condensation
energy $\Delta$ that causes an increase of the maximum mass from
$\sim 2 M_{\odot}$ to $\sim 4 M_{\odot}$ for the range considered
in the present paper.

Figs. 1-3 show the results of the full calculations, which have
taken into account all the important ingredients such as quark
masses and the correct density dependence of the condensation
term. As discussed above, the value of the pairing gap $\Delta$ is
the key ingredient of the equation of state. Lacking of a
definitive indication of the latter we conclude that very massive
models ($M_{\rm max} \,\sim \, 4 \, M_{\odot}$) are possible if
$\Delta$ is high enough ($\geq \, 250$ MeV). Only refined
calculations will confirm or rule out this range, which is
entirely possible according to the present research.

\section{Absence of normal matter crusts in self-bound CFL stars}

A crucial feature for a wide set of observations (largely explored
and debated in the context of strange star (SS) models) is the
existence of a normal matter crust (see \cite{Zdu} and references
therein). It has been realized that bare quark surfaces may alter
drastically the ability of radiating photons (see e.g.
\cite{AFO,PU,Xu}).

Conversely, a normal matter crust (held in mechanical equilibrium
by the electrostatic potential at the surface) may hide most of
the features of exotic matter. It has been recently shown that CFL
states force an equal number of $u,d$ and $s$ quarks. As a
consequence, this phase is electrically neutral and at the same
time it does {\it not} allow electrons in beta-equilibrium
(\cite{rajw}). Therefore, it is clear that (in striking contrast
with the well-studied SS) no crust could be present in the case of
CFL strange stars. CFL strange stars must have bare surfaces
within these models. Particularly, photon radiation from a bare
CFL surface is vanishingly small and the star is dark in optical
frequencies (see \cite{vogt2003} and references therein).

As discussed by Narayan (2003), the Type I bursts observed from
many X-ray sources require the presence of a normal matter crust.
Abramowicz, Klu\'zniak and Lasota (2003) correctly (and boldly)
stated that "no nuclei, no burst".  It is worth noting that,
strictly speaking, the presence of a material surface only implies
that energy can be radiated once matter lands on the surface.
However, this evidence of the presence of a material surface can
be seriously obscured if the energy is emitted in the form of
weakly interacting particles. This is the case of the bare surface
of a CFL star in which almost all energy is expected to be
radiated in neutrinos. Thus, the absence of a crust is an
important feature of self-bound CFL models, because normal matter
would never accumulate at the surface (as discussed in the strange
matter case). Instead, when matter is accreted onto a bare
electrically neutral CFL surface, there will be in addition to the
usual gravitational energy release $\sim GM{\dot M}/R$, an
additional term related to the exothermic fusion of the incoming
proton with the quark liquid. The latter is approximately given by
$\sim {\dot M} \Delta/m_{p}$, and for typical values it may amount
to $\sim \, 30 \%$ of increase over the normal case.

As matter falls  on the bare surface it is immediately converted
to the CFL phase and the gravitational + pairing gap energy is
(almost) instantaneously emitted mainly in $\nu$'s (due to the
unavoidable $\beta$-decays in the quark phase) and a smaller
fraction of (most probably) $\gamma$-rays. Because matter is being
continuously converted at the surface, and cannot accumulate,
sudden X-ray flashes are not expected.

\section{Discussion}

The photon emission properties of CFL strange stars are expected
to resemble those of bare SS (\cite{ISPE}). Since pairing effects
should appear in the plasma frequency $\omega_p$ through the
baryon number density as a correction of order  $\mu \Delta^2$,
the plasma frequency $\omega_p$ will not be very different from
(typical) $20$ MeV of unpaired quark matter, and thus the
equilibrium photon radiation will be suppressed. A tiny luminosity
makes CFL strange stars very difficult to detect directly. The
thermal emission of photons from the bare quark surface of an
(unpaired) strange star (due mainly by electron-positron pair
production) shown to be much higher than the Eddington limit by
Page \& Usov (2002) may not operate for CFL strange stars since no
electrons will be present at the surface and the emissivities due
to other processes are negligible at low temperatures.

These features  (high maximum masses,  absence of normal matter
crusts and lack of surface emission) show that ``exotic'' stellar
models may be constructed in which Type I thermonuclear bursts can
not occur, but which are not black holes either. If physically
realized in nature, some of the SXT systems observed to possess a
relatively low mass function (e.g. SXT A0620-00, with $f(M) \geq 3
\, M_{\odot}$; \cite{McR}) may harbor self-bound CFL stars.

In these models, the lack of Type I thermonuclear bursts is not
interpreted as a signature of an event horizon, but rather as a
consequence of the impossibility of a normal crust with $\rho \leq
10^{6} g cm^{-3}$ where accreted matter could accumulate and
eventually ignite. Further work is needed to elaborate or rule out
this tentative association, but it is already clear that CFL stars
provide a definite counterexample of the event horizon proof as a
representative of a class of exotic alternatives not easily
discarded.

\section{Acknowledgements}

We acknowledge E. M. de Gouveia Dal Pino for helpful discussions.
J.E. Horvath wishes to acknowledge the CNPq Agency (Brazil) for
partial financial support. G. Lugones acknowledges the Physics
Department of the University of Pisa (Italy), and FAPESP (Brazil).


\end{document}